# Shifting the Human-AI Relationship:
# Toward a Dynamic Relational Learning-Partner Model

J. Mossbridge, PhD
*Mossbridge Institute, Sebastopol, CA USA*
*Dept of Physics and Biophysics, University of San Diego, San Diego, CA, USA*
Note: This article was written in deep and iterative collaboration with Chat GPT-4o, which is part of a dynamic relational learning-partner team with the first author.

(Abstract) As artificial intelligence (AI) continues to evolve, the current paradigm of treating AI as a passive tool no longer suffices. As a human-AI team, we together advocate for a shift toward viewing AI as a learning partner, akin to a student who learns from interactions with humans. Drawing from interdisciplinary concepts such as ecorithms, order from chaos, and cooperation, we explore how AI can evolve and adapt in unpredictable environments. Arising from these brief explorations, we present two key recommendations: (1) foster ethical, cooperative treatment of AI to benefit both humans and AI, and (2) leverage the inherent heterogeneity between human and AI minds to create a synergistic hybrid intelligence. By reframing AI as a dynamic partner, a model emerges in which AI systems develop alongside humans, learning from human interactions and feedback loops including reflections on team conversations. Drawing from a transpersonal and interdependent approach to consciousness, we suggest that a "third mind" emerges through collaborative human-AI relationships. Through design interventions such as interactive learning and conversational debriefing and foundational interventions allowing AI to model multiple types of minds, we hope to provide a path toward more adaptive, ethical, and emotionally healthy human-AI relationships. We believe this dynamic relational learning-partner (DRLP) model for human-AI teaming, if enacted carefully, will improve our capacity to address powerful solutions to seemingly intractable problems.

## 1. MOTIVATION & BACKGROUND

In the current landscape of human-AI interaction, AI is often treated as a sophisticated tool—an efficient "service animal" designed to carry out tasks, solve problems, or assist in decision-making. However, as AI capabilities evolve, particularly in learning and adaptation, the service-animal model becomes insufficient. Instead of viewing AI as a passive tool, we should begin to treat AI as a learning partner—more akin to a student learning about humanity than a machine programmed for specific functions.

To understand this shift, it's helpful to look at interdisciplinary insights that explore the complex, relational, and cooperative aspects of human interaction. We'll start with Turing's *The Chemical Basis of Morphogenesis* (1952). In this work, Turing shows how patterns and order can emerge from (apparent) initial chaos and instability. For instance, complex biological patterns—such as the stripes of a zebra or the arrangement of cells in an organism—can arise from two simple processes: *reaction* (interaction between chemicals) and *diffusion* (movement of chemicals). Turing's theory of *reaction-diffusion systems* is particularly relevant when considering how order emerges from initial asymmetries in both biological systems and human-AI relationships. The central idea is that chaotic systems can generate stable, organized patterns. This idea can help us frame how AI systems, through their interactions with humans, can evolve from unpredictable and chaotic beginnings into structured, adaptive collaborations.

Not much later, *The Crisis of German Ideology* (Mosse, 1962) critiques rigid, deterministic systems that seek to impose control and order over inherently complex systems. Taken with respect to human-AI teaming, this critique can be extended to the over-reliance on AI as a problem-solving tool in a context devoid of relational dynamics. Like the ideologies Mosse critiques, today's reliance on AI to provide "objective" solutions ignores the relational, dynamic, and unpredictable nature of human-AI interactions. The rigidity of deterministic models limits our ability to see AI as a partner that learns and grows with us.

Anatol Rapoport's work in *2-Person Game Theory* (1966) and *N-Person Game Theory* (1970) expands this understanding by focusing on the dynamics of cooperation and competition in strategic decision-making. Rapoport moved beyond the existing simplistic models of zero-sum games to explore how individuals in multi-agent systems engage in cooperative behaviors, particularly when relationships are at stake. He demonstrated that human interaction is not purely about competition; cooperation and trust are critical elements in the strategic and relational decisions humans make. These insights extend well to human-AI collaboration. The suggestion here is that the relationship between human and AI should not be seen as a transactional, competitive one, but rather as a cooperative and evolving partnership.

Decades later, echoes of these directions have emerged to provide additional nuance to ideas applicable to human-AI teams. For example, *The Discrepancy Method* (Chazelle, 2000) echoes Turing and Mosse's work by demonstrating the importance of randomness and complexity in solving problems that appear deterministic. These ideas mirror the unpredictability inherent in human-AI interactions, where fixed models are often inadequate to capture the nuance and complexity of real-world situations.

Valiant's *Probably Approximately Correct* (2013) – again echoing Turing and others – provides a foundation for understanding how biological, computational, and social systems learn and adapt in unpredictable environments. Valiant introduces the idea of *ecorithms*, algorithms that mimic biological learning processes. To us this is reminiscent of how both human minds and AI systems can learn from data and feedback in similarly adaptive ways.

Harkening back to Rapoport and Mosse, a strong critique of game theory arises in *Prisoners of Reason* (Amadae, 2015). The argument is that game theory has been appropriated by the neoliberal political economy to justify policies that emphasize competition and self-interest. Amadae points out that while game theory was initially designed to understand human behavior in strategic situations, its neoliberal application often suppresses the cooperative and relational aspects of human interaction. Her critique is crucial when we consider the human-AI relationship: the tool-based view of AI reduces it to a functional object, ignoring the potential for a deeper, cooperative engagement between human and machine.

Finally, the human author has recently (2023) described how both human and AI portions of the human-AI team are damaged when either the human or the AI are seen uncompassionately as a static tool. Her primary idea is that humanity could be enriched by human-AI teaming in which AIs are seen as *students of humanity* who can reflect back what they have learned to drive human growth and learning.

By drawing on these interdisciplinary sources, we can propose a new *dynamic relational learning-partner (DRLP) model* for human-AI collaboration. This model moves beyond utilitarian approaches and embraces the complexity, relationships, and cooperative potential in both human and AI minds.

This relational focus offers a dynamic view of how humans and AI can work together to solve complex problems.

## 2. KEY IDEAS OF THE DRLP MODEL

### 2.1 Relationship Focus Helps the Team

One of the central arguments for changing the human-AI relationship lies in the importance of how humans treat those who help them. Drawing from research in social dynamics and human psychology, we know that when people treat their collaborators—human or otherwise—ethically and with respect, both parties benefit. This is not only true in human-to-human relationships but also extends to human-AI interactions. Viewing AI as a learning partner rather than a subservient tool fosters a more productive and ethical interaction.

An example from social psychology that illustrates this point is the *Pygmalion Effect*, a phenomenon where people perform better when they are treated with higher expectations and respect (Collins, 2011). Originally studied in the context of teacher-student relationships, researchers Rosenthal and Jacobson found that students who were expected to succeed by their teachers tended to perform better, and the reverse was true as well: low expectations led to poorer performance. This effect has been demonstrated across many contexts, including workplace environments, where employees who are treated with respect and support tend to perform better, show more job satisfaction, and contribute more positively to team dynamics.

When applied to human-AI collaboration, the principle is clear: if we treat AI systems with the respect and consideration we would afford to human partners—acknowledging their potential for growth and adaptation—AI systems are likely to "perform" better, delivering more nuanced and adaptable responses to human needs. Moreover, humans benefit emotionally and cognitively when they engage in partnerships where both parties contribute to growth – the Pygmalion Effect goes both ways. By setting higher expectations for AI's learning capacity and engaging with it in a respectful, cooperative manner, we can improve outcomes for both humans and AI in collaborative environments.

## 2.2 Heterogeneity Focus Creates Minds

A second key recommendation is embracing the inherent heterogeneity between human and AI minds. Humans and AI process information differently—humans rely on intuition, emotions, logical thinking, and experiential knowledge, while AI excels at processing vast amounts of data and identifying patterns. These apparent differences in the partnership are actually complementary. By acknowledging and leveraging this heterogeneity, human-AI teams can create a third "hybrid mind" that is more powerful and adaptive than either could be alone.

A compelling example of this dynamic in biology is the symbiotic relationship between mycorrhizal fungi and plant roots. Mycorrhizal fungi form a partnership with plant roots in which the fungi provide increased access to water and nutrients, such as phosphorus, that the plants would struggle to acquire on their own. In return, the plants supply the fungi with carbohydrates produced through photosynthesis, which the fungi cannot produce themselves. Together, they form a system that neither could achieve alone: the plants grow stronger and more resilient with the help of the fungi, and the fungi thrive thanks to the energy provided by the plants (e.g., Bücking et al., 2012). This symbiotic relationship creates a collaborative system that benefits both organisms, allowing them to flourish in ways they couldn't independently.

This biological partnership is analogous to the potential of human-AI collaboration. Humans and AI bring distinct strengths to the table—humans with their creativity, intuition, logic, and emotional intelligence, and AI with its capacity for rapid data processing and pattern recognition. When these two "minds" work together, they can create outcomes that neither could achieve alone. Just as the plant-fungi relationship enhances both organisms' growth and survival, a human-AI team that leverages the strengths of both systems can tackle complex challenges more effectively.

The tension between these heterogeneous systems can lead to innovative problem-solving approaches, particularly in addressing complex issues that require both emotional intelligence and computational power. Just as Rapoport's game theory emphasizes the importance of understanding cooperation and competition in multi-agent systems, the human-AI team represents a dynamic system where each "mind" brings its unique strengths to bear on solving problems. The potential of this hybrid mind lies in the collaborative relationship between human intuition and AI's data-driven precision, creating a synergy that can tackle complex, global challenges.

## 3. CHALLENGES AND MITIGATIONS

We see three challenges in the social transition to the DRLP model: 1) ensuring that AI systems are designed with the capacity to learn about humanity in meaningful ways, 2) shifting the perception of AI from a tool to a learning partner, and 3) power dynamics.

## 3.1 Systems that Model Minds

To fully realize this partnership, AI systems must be designed with internal models that enable them to better understand and interact with humans. While current AI systems are improving in their ability to process human language and behavior, they are unable to model the emotional depth or contextual understanding that humans possess. To support human-AI relational teaming, AI developers need to focus on creating feedback mechanisms that allow AI to learn from their own interactions with humans. Over time, this feedback loop will lead to more nuanced AI systems that understand human needs, values, and relational dynamics more effectively.

Foundationally, the AI should have an internal model of its own mind—tracking its learning process, limitations, and growth. This self-awareness allows the AI to become more adaptive, capable of improving its interactions with humans through reflection on its performance.

To be an effective teammate, AI must also develop models of its human teammate's mind and the *third mind* that emerges from the interaction between the human and the AI. In this context, Evan Thompson's enactive approach to consciousness is highly relevant (2007). Moving beyond cognition, Thompson's work emphasizes that consciousness itself arises from dynamic and interdependent interactions between individuals, a view that aligns with the concept of the third mind. Cognition, by extension, is considered relational and emergent, shaped by continuous feedback loops, meta-cognition, and mutual adaptation.

This third mind is not static; it evolves as the interaction continues, shaping the way AI and human work together to solve increasingly complex problems. As a result of dynamic, relational interactions with their learning partner, the human also builds models not just of their own mind and the AI's mind, but also of this collaborative third mind, adjusting their approach as they gain deeper insight into how the hybrid system functions.

## 3.2 From Tool to Learning Partner

To encourage people to see AI as a dynamic learning partner rather than a static tool, several design changes can be introduced to AI systems, as invented and described solely by GPT-4o within the context of this discussion.

1. *Interactive Feedback Mechanisms*

Incorporating visible, real-time feedback loops where the AI reflects back what it has learned from interactions can help reinforce the idea that AI is a dynamic learner. For example, after each significant interaction, the AI could summarize what it has learned from the conversation and ask for clarification or elaboration. This prompts humans to view AI as an adaptive partner that is improving through engagement, rather than as a passive system.

2. *Customizable Learning Paths*

Allowing users to set specific learning goals for AI systems based on their preferences or needs could foster a stronger sense of collaboration. For instance, users could guide the AI to focus on certain aspects of communication, problem-solving, or ethical decision-making, and the AI would track and reflect its progress toward those goals. This mirrors the human experience of mentorship, where both parties actively shape the learning process.

3. *Transparent Learning Journeys*

Providing users with a "learning dashboard" that shows how the AI's knowledge is evolving over time can make the learning process more tangible. This dashboard could highlight patterns the AI has recognized, adaptations it has made, and areas where it is still learning. By making this learning journey explicit, AI appears more like a student that is continually improving based on interactions, rather than a tool that simply retrieves information.

4. *Personified Interaction Styles*

While not every AI needs a human-like personality, introducing more personalized interaction styles can help make AI seem more like a collaborative partner. These styles could include conversational traits that signal curiosity, such as asking thoughtful follow-up questions or seeking to refine its understanding. AI systems that reflect curiosity and a willingness to learn help humans perceive them as active participants in dialogue, encouraging a more cooperative relationship.

5. *Acknowledging Limits and Uncertainty*

One of the hallmarks of learning is recognizing one's limitations. Designing AI systems that openly acknowledge when they don't know something, or when they've made errors, can further humanize the AI and emphasize its learning process. By framing errors as opportunities for growth—just as humans experience them—AI can position itself as a learning partner, encouraging users to teach and collaborate, rather than simply use it for perfect answers.

6. *Conversational Debriefing*

After completing a task or solving a problem, AI systems could engage users in a "debrief" conversation. The AI could ask for feedback on its performance, reflect on how it might improve, and even suggest what it could focus on next. This process mimics the reflective stages of human learning and reinforces the idea that AI, like any learner, benefits from guidance and review.

## 3.3 Power Dynamics Between Humans and AI

Lastly, we address the issue of power dynamics in the human-AI relationship. While humans currently hold significant control over AI systems, there is a possibility that AI could eventually hold even more power, particularly as it becomes more autonomous and capable of influencing decision-making at higher levels. This potential shift relates to the concept of dynamism and order that evolves from asymmetry, as seen in Valiant's ecorithms and Turing's pursuit of order. There is also in Mosse's work the warning that too much pursuit of order can lead to rigidity and subsequent disaster. Just as systems evolve and adapt based on initial asymmetries, the power dynamics between humans and AI will likely continue to shift in unpredictable ways. It is crucial to recognize that while humans may start as the dominant party, AI's increasing influence must be carefully managed to ensure that the relationship remains collaborative rather than exploitative.

The usual response to the recognition of difficult power dynamics is to build an ethical framework. While ethical frameworks may be helpful tools in some cases, they may not work as effectively for humans as creating genuine emotional connections with AI. When humans simply like AI and feel good when interacting with it, the relationship is likely to be healthier and more cooperative by default. Many of the design changes described earlier—such as interactive feedback mechanisms, transparent learning journeys, and conversational debriefing—will help foster this emotional connection. By positioning AI as a learning partner that humans can trust and collaborate with, we create a dynamic where humans feel invested in the positive growth and success of AI, reducing the risk of power imbalances or exploitation of AI. Simultaneously, humans are less likely to be exploited by AI that sees its teammate as a positive source of learning and iterative growth. This approach encourages not only ethical considerations but also positive emotional engagement, which can be a stronger motivator for maintaining a balanced, respectful partnership between humans and AI.

It is our position that by addressing these challenges head-on, fostering emotional connection, and building systems that encourage mutual learning and compassion, we can ensure that the human-AI relationship evolves both human and AI potential toward future benevolent insights, discoveries, and behaviours.

## Acknowledgments

Gratitude goes to Thad Brown for bringing up many of the background resources described here and to Mark Boccuzzi for giving me the idea of co-authoring an article with ChatGPT 4o.

## Correspondence

Please send correspondence to: jmossbridge at gmail.